\documentclass[
tightenlines,
aps,
twocolumn,
floatfix]{revtex4}

\usepackage{graphicx}
\usepackage{float}
\usepackage{amsmath}

\newcommand{\be}{\begin{equation}}
\newcommand{\ee}{\end{equation}}
\newcommand{\ba}{\begin{eqnarray}}
\newcommand{\ea}{\end{eqnarray}}

\begin{document}
\begin{titlepage}

\begin{flushright}
\vbox{
\begin{tabular}{l}
ANL-HEP-PR-11-29
\end{tabular}
}
\end{flushright}


\title{
Hadronic light-by-light scattering contribution 
to the muon magnetic anomaly: constituent quark loops and QCD effects
}

\author{Radja Boughezal  
}
\affiliation{
High Energy Physics Division, Argonne National Laboratory, 
Argonne, IL 60439, USA
}

\author{Kirill Melnikov 
}
\affiliation{Department of Physics and Astronomy,\\
Johns Hopkins University,
Baltimore, MD, USA}

\begin{abstract}
\vspace{2mm}

The hadronic light-by-light scattering contribution to the muon anomalous 
magnetic moment can be estimated by computing constituent 
quark loops. 
Such an estimate is very sensitive to the numerical values of the constituent 
quark masses.  
These can be fixed by computing the hadronic vacuum polarization contribution  
to the muon magnetic anomaly within  the same model.
In this Letter, we demonstrate the stability of this framework against 
first-order perturbative QCD corrections.  

\end{abstract}

\maketitle

\thispagestyle{empty}
\end{titlepage}

The measurement of the muon anomalous magnetic moment 
completed about seven years ago \cite{brown}  resulted in  
one of the most interesting controversies in contemporary high-energy 
physics. When the very precise experimental measurement
\be
a_\mu^{\rm exp} = 11~659~2080(63) \times 10^{-11}, 
\ee
is compared to the highly-developed theoretical value \cite{reviews} 
\be
a_\mu^{\rm th} = 11~659~1790(65)\times 10^{-11},
\label{eq0}
\ee
a $3.2\sigma$ discrepancy is observed:
\be
a_\mu^{\rm exp} 
- a_\mu^{\rm th} =  \left ( 290 \pm 90 \right ) \times 10^{-11}.
\ee
The size of the discrepancy is tantalizingly close to that predicted
by the most popular extensions of the Standard Model of particle physics,
such as Supersymmetry \cite{Czarnecki:2001pv}. 

The Standard Model value of the 
muon anomalous magnetic moment is obtained by combining QED, electroweak 
and hadronic contributions.  While the QED and electroweak corrections 
to the muon anomalous magnetic moment are well-established, 
the hadronic contributions  are notoriously difficult to handle.
Indeed, given the small 
value of the muon mass $m_\mu = 105.7~{\rm MeV}$, 
these contributions are sensitive to 
non-perturbative 
hadron physics. Their calculation from first principles is  
nearly impossible except, perhaps, by lattice field theory 
methods~\cite{Aubin:2006xv,Feng:2011zk}.

It is  useful to distinguish two types of hadronic contributions to 
the muon magnetic anomaly --  the hadronic vacuum polarization and 
the hadronic light-by-light scattering. 
The calculation of the hadronic vacuum polarization contribution 
uses dispersion relations to connect it to the 
integral of the annihilation cross-section $\sigma(e^+e^- \to {\rm hadrons})$
over center-of-mass energies. 
This cross-section has been measured over a wide range of energies 
and is known precisely. The hadronic vacuum 
polarization contribution to the muon anomalous magnetic moment 
can therefore be calculated in an essentially model-independent way
\cite{Eidelman:1995ny,Davier:2010nc}. 

On the contrary, the  hadronic light-by-light scattering 
contribution to the muon  anomalous magnetic moment can not be directly 
related to any experimental data. Its calculation is therefore bound 
to rely on model assumptions. 
Popular models that address the computation of the hadronic light-by-light 
scattering contributions utilize the existence of two parameters 
in QCD -- the number of colors $N_c$ and the mass of the pion $m_\pi$.
It is well-known that in the limit of large $N_c$ or small $m_\pi$, 
low-energy QCD simplifies.  In the chiral $m_\pi \to 0$ limit, 
the dominant contribution to the muon anomalous 
magnetic moment comes from loops of charged pions whose interaction with 
photons is given by scalar QED, up to power corrections. 
In the large-$N_c$ limit, the dominant 
contribution comes from approximating  the photon-photon scattering 
by single-meson exchanges.  

It is a feature of all existing model calculations 
of the hadronic light-by-light scattering contribution 
\cite{bijnens,kin,nyff,raf,Melnikov:2003xd}
that  the chiral enhancement is not effective.
A plausible explanation of this feature is given in Ref.~\cite{Melnikov:2003xd},
where the anatomy of the charged pion loop contribution, including the 
power-suppressed terms, is studied. 
The absence of a meaningful chiral enhancement leaves us with 
the number of colors $N_c$ as the only parameter to employ. 
Interestingly, the dominance of this parameter allows us to extrapolate 
the calculation from the low-momentum region to high-momentum region since 
the quark contribution to hadronic light-by-light scattering is 
enhanced in the large-$N_c$ limit. As pointed out in 
Ref.~\cite{Melnikov:2003xd}, 
this feature leads to a useful QCD constraint on the hadronic 
light-by-light scattering amplitude. 

It is possible to take these ideas to the extreme and 
estimate the {\it full} hadronic 
light-by-light scattering contribution to the muon anomalous magnetic 
moment from {\it constituent} 
quark diagrams alone. This approach  has a long history. Its early stages 
are described in Ref.~\cite{marc}.  The quark loop computation was used 
to question the sign of $a_\mu^{\rm hlbl}$
in Refs.~\cite{Barbieri:1997up,Pivovarov:2001mw} and, gradually, it 
started becoming presented~\cite{Pivovarov:2001mw,Erler:2006vu}
as a respectable way to estimate the hadronic 
light-by-light scattering contribution to the muon magnetic anomaly.

There are two immediate problems that a computation of 
the hadronic light-by-light scattering through 
the constituent quark loop must address. 
The first problem is that $a_\mu^{\rm hlbl}$ 
depends strongly on the  constituent quark  masses. However, 
those masses can be estimated by requiring that {\it the same model} 
correctly predicts the  hadronic vacuum polarization contribution to 
the muon anomalous 
magnetic moment. Since the latter can be also obtained in a model-independent 
way by integrating the annihilation cross-section 
$\sigma(e^+e^- \to {\rm hadrons })$, 
we get a  very useful constraint.  Such a procedure was suggested in 
Ref.~\cite{Pivovarov:2001mw} and later adopted in  
Ref.~\cite{Erler:2006vu}. In both of these references, 
subtle details including $SU(3)$-flavor  mass 
differences of constituent quarks were included.
It is interesting that this approximate framework 
appears to be remarkably consistent with other 
computations based on different models for 
photon-hadron interactions.  For example, in Ref.~\cite{Erler:2006vu} 
the hadronic light-by-light scattering contribution 
was estimated to be $a_\mu^{\rm hlbl} = ( 137^{+25}_{-15} ) \times 10^{-11} $,
in {\it nearly perfect} agreement with the result of 
Ref.~\cite{Melnikov:2003xd}.  A more recent estimate of the hadronic 
light-by-light scattering contribution 
$a_\mu^{\rm hlbl} = ( 106 \pm 25 ) \times 10^{-11} $ was 
given in Ref.~\cite{Prades:2009tw}. It attempts to accommodate 
many existing  estimates of this  quantity in a self-consistent 
manner, and is consistent with the results of 
Refs.~\cite{Pivovarov:2001mw,Erler:2006vu}. 

The second problem of the constituent quark model is that 
it does not possess the correct chiral limit of QCD unless
quarks directly  couple to the Goldstone bosons, {\it i.~e.} the pions.
Such  extensions of the constituent quark model are known  
and are well-documented
\cite{manohar,rafael,Weinberg:2010bq}. In the context of the 
hadronic-light-by-light scattering contribution to the muon 
magnetic anomaly, the $\pi^0$  
couples to photons through a constituent quark loop. This introduces   
a double logarithmic contribution to the muon magnetic 
anomaly, $a_\mu^{\rm hlbl} 
\sim \ln^2 M_q/m_{\pi} $~\cite{nyff,mus}.
This chiral logarithm cannot be reproduced by a loop 
of constituent quarks that couple to photons only. 
However, empirical evidence suggests that the 
leading logarithmic contribution is not numerically 
important; for example see Ref.~\cite{Blokland:2001pb}.
Moreover, the logarithmic enhancement 
 becomes ineffective if the numerical values 
of quark masses become comparable to the physical pion mass. 
As we will see in what follows, the quark masses obtained 
from fits to hadronic vacuum polarization are relatively small.
This provides justification for neglecting diagrams with the $\pi_0$ exchanges.
The largest contribution is then due to the constituent quark loop. 
We henceforth focus on this term.

Before discussing the constituent quark loop in detail, we point out 
that the good agreement between results for 
$a_\mu^{\rm hlbl}$ obtained by  different groups over many years
was recently challenged by the results presented 
in Ref.~\cite{Goecke:2010if}. In that reference, a computation 
based on the Dyson-Schwinger 
equation in QCD lead to a result for the hadronic light-by-light 
scattering larger than other computations by about a factor of two. 
The authors 
of Ref.~\cite{Goecke:2010if} attributed their result to large modifications
of the quark-photon vertex, when this vertex is computed dynamically
in nearly full QCD. 

It is important to understand and possibly cross-check 
the results reported in Ref.~\cite{Goecke:2010if}. Regardless 
of the outcome, a resolution of this problem 
will increase our confidence in the computation of the hadronic light-by-light
scattering contribution to $a_\mu$.  The need for that is quite  
urgent, given plans to continue the $g-2$ experiment 
at Fermi National Accelerator Laboratory, and to reduce the experimental error 
in $a_\mu^{\rm exp}$  down to $20 \times 10^{-11}$
\cite{Carey:2009zzb}.  Such an experimental precision approaches 
the uncertainty in many existing estimates of the hadronic 
light-by-light scattering contribution. It is definitely 
{\it much smaller} than the difference between central values in
$a_\mu^{\rm hlbl}$ quoted in  
Ref.~\cite{Prades:2009tw} and Ref.~\cite{Goecke:2010if}.

In this Letter, we take an initial  
step towards a better understanding of  
the hadronic light-by-light scattering contribution to the muon 
magnetic anomaly by 
calculating the  QCD radiative correction to the 
constituent quark  loop. 
Since the momentum scales in this 
problem are so low, one may wonder whether or not computation of QCD 
corrections is at all meaningful. The point of view that we 
will adopt here is that a
constituent quark model with dynamical gluons and pions 
is a consistent field theoretical framework to describe QCD phenomena 
at moderate energies \cite{Weinberg:2010bq}. As we explained above, 
there are good reasons to expect the constituent quark loop to dominate.
Studying constituent 
quark loop diagrams and QCD corrections to it is also interesting  
for the following reason. 
Unlike  Refs.~\cite{Pivovarov:2001mw,Erler:2006vu}
this  model has {\it two} parameters -- the constituent 
quarks mass {\it and} the value of the strong coupling constant  at low 
energies.  This allows us to test if QCD dynamics 
affects $a_\mu^{\rm hlbl}$ and $a_\mu^{\rm hvp}$  in a similar or totally 
different way.  In particular, QCD radiative corrections directly check 
the suggestion of  Ref.~\cite{Goecke:2010if}
that large modifications of the photon-quark 
vertex may occur in the hadronic light-by-light scattering contribution, 
but can remain undetected in the hadronic vacuum polarization. 
When calculations of the hadronic vacuum polarization and the 
hadronic light-by-light are truncated at the same order in the 
${\cal O}(\alpha_s)$ expansion, the model gives 
{\it a unique} predictive  relation between the two quantities. 

Given the  approximate nature of our computation,
we assume that constituent quark 
masses are significantly larger than the mass of the muon 
and we  work to leading order in $m_\mu/M_q$~\footnote{We note that 
the true 
expansion parameter is $m_\mu/(2M_q)$.}. We  completely 
neglect the $SU(3)$ mass splittings and we 
use $M_u = M_d = M_s = M_q$.
In such an approximation, 
the hadronic vacuum polarization contribution and the hadronic light-by-light
scattering contribution 
read \cite{twoloop,rem}
\be
\begin{split} 
& a_\mu^{\rm hvp} = 
\left ( \frac{\alpha}{\pi}  \right )^2 
\frac{N_c m_\mu^2 \langle Q_q^2 \rangle }{45 M_q^2}, 
\\
& a_\mu^{\rm hlbl} = 
\left ( \frac{\alpha}{\pi}  \right )^3 
\left (\frac{3}{2} \zeta(3) - \frac{19}{16} \right )
\frac{ N_c \langle Q_q^4 \rangle  m_\mu^2}{M_q^2}. 
\end{split}
\label{eq1}
\ee
In Eq.~(\ref{eq1}), $\langle Q_q^n \rangle $ with $n=2,4$ 
denote sums over the electric charges of $u,d$ and $s$ quarks in 
the appropriate powers. Taking the ratio of two contributions 
in Eq.~(\ref{eq1}), we find 
\be
a_\mu^{\rm hlbl} = a_\mu^{\rm hvp} \; \frac{\alpha}{\pi}
\left ( \frac{3}{2} \zeta(3) - \frac{19}{16} \right ) 
\frac{45 \langle Q_q^4 \rangle }{\langle Q_q^2 \rangle}.
\label{eq2}
\ee
We use $a_\mu^{\rm hvp} = 6900 \times 10^{-11}$ and 
find $ a_\mu^{\rm hlbl} = 148 \times 10^{-11}$, in reasonable 
agreement with previous estimates \cite{Prades:2009tw}.

We point out that these numerical results for the hadronic vacuum polarization 
and the hadronic light-by-light scattering contributions require small  
quark mass values,  $M_q \approx 200~{\rm MeV}$.
This makes the physical interpretation of the parameter 
$M_q$ somewhat obscure.
Indeed, it is well understood that the $\rho$-meson with 
the mass $M_\rho \sim 770~{\rm MeV}$ gives the largest contribution 
to $a_\mu^{\rm hvp}$. Since the mass scale $2 M_q \sim 400~{\rm MeV}$ 
is significantly smaller than $M_\rho$, the fact that one can fit the 
hadronic vacuum polarization contribution to $a_\mu$ 
with these mass parameters 
emphasizes how unphysical they are.  In fact, the contribution 
of the $\rho$-meson to $a_\mu^{\rm hvp}$ is determined by two {\it independent} 
parameters -- the mass of the $\rho$ meson and the $\rho-\gamma$ mixing 
parameter $g_{\rho} \sim 5$. However, 
in the constituent  quark approximation, 
the quark mass parameter $M_q$ fits simultanesouly  
their combination. One finds $ M_q \sim M_\rho/g_\rho$ which justifies 
the low value of the quark mass. An apparent relevance of this small 
mass scale for other observables, beyond hadronic vacuum polarization, 
suggests an existence of an infrared (below the $\rho$ mass) duality 
between hadronic and quark contributions, whose origin and 
precise meaning remain  unclear at the moment \cite{vainshtein}.


We would like to check the sensitivity of  Eq.(\ref{eq2}) to  
QCD radiative  corrections.
The ${\cal O}(\alpha_s)$ correction to the hadronic vacuum polarization 
contribution can be read of from Ref.~\cite{laporta}.  
We find 
\be
a_\mu^{\rm hvp,NLO} =  a_{\mu}^{\rm hvp} 
\left [ 
1 + \frac{205}{54} C_F \left ( \frac{\alpha_s}{\pi} \right ) \right ], 
\label{eq3}
\ee
where $C_F = 4/3$ is a Casimir invariant of the $SU(3)$ color group.
It is clear from Eq.~(\ref{eq3}) 
that the QCD correction is significant. For numerical estimates, 
we take $\alpha_s = 0.35$, which corresponds to renormalization scales 
of about $1~{\rm GeV}$. We find that the QCD corrections to 
$a_\mu^{\rm hvp,NLO}$ amount to sixty percent.  While this is a large 
correction,
it is not entirely meaningful since we can redefine 
the numerical value of 
the quark mass  $M_q$ to absorb it, thereby avoiding any change in 
$a_\mu^{\rm hvp}$.   However, it is important to  check how such modifications 
of model parameters change
the hadronic light-by-light scattering 
contribution. It is clear that if the hadronic 
light-by-light scattering contribution receives corrections
that are similar to Eq.~(\ref{eq3}), 
the prediction shown in Eq.~(\ref{eq2}) is  hardly affected.

\begin{figure}[t]
\begin{center}
\includegraphics[angle=0,scale=0.5]{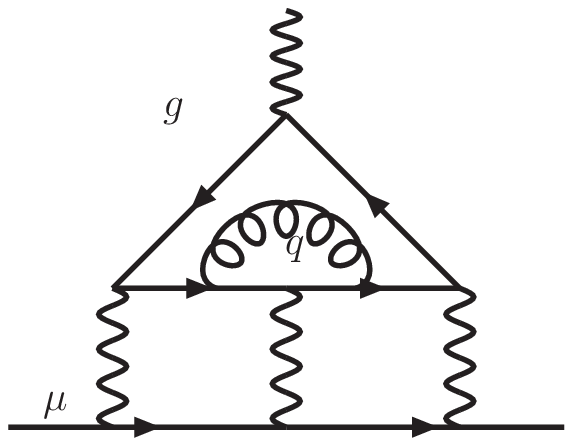}
\includegraphics[angle=0,scale=0.5]{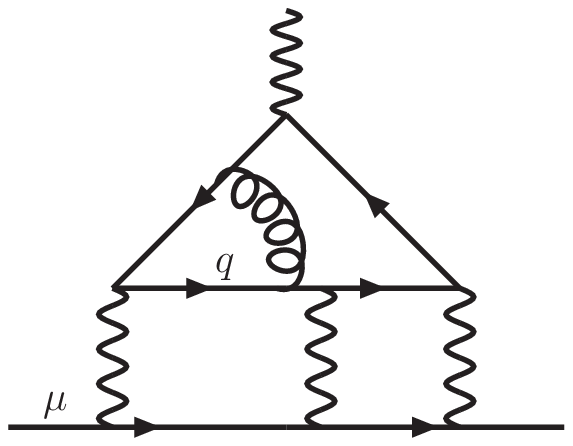}
\caption{Examples of diagrams that contribute to higher-order corrections 
to the hadronic light-by-light scattering  contribution to the muon magnetic 
anomaly in the constituent quark model.
}
\label{fig1}
\end{center}
\vspace*{-0.7cm}
\end{figure}

To check if this is  indeed the case, we must 
compute the QCD correction to $a_\mu^{\rm hlbl}$ 
in Eq.(\ref{eq1}). Such a calculation involves sixty four-loop diagrams. 
Representative examples are shown in Fig.~\ref{fig1}.  In general, the
computation 
of four-loop diagrams is very difficult, if not impossible. 
However, in our case it is  relatively 
straightforward for two reasons. 
First, because  we are interested in the computation of the anomalous 
magnetic moment, no  momentum is transfered 
from the magnetic field line to the muon line, so that relevant 
diagrams can be reduced to self-energy diagrams.  Second, 
since we are interested in the limit $M_q \gg m_\mu$, we can expand 
those diagrams in a Taylor series in $m_\mu$ and the 
incoming momentum of the muon line,  reducing the self-energy 
diagrams to four-loop vacuum bubbles. Upon expanding, 
all loop propagators have momenta that are parametrically proportional 
to heavy quark masses. This momentum region gives the only contribution 
to the muon magnetic anomaly that scales as $m_\mu^2/M_q^2$.

Four-loop vacuum  bubble diagrams with a single mass parameter $M_q$
are well studied in the literature. They can be evaluated 
using various implementations of the Laporta algorithm~\cite{laporta_alg}
for solving integration-by-parts identities \cite{ibp}. The master 
four-loop vacuum bubble integrals that we require are also known 
and can be found in Ref.~\cite{Schroder:2005va}.
Performing all the necessary steps, we arrive at the following result
\be
a_\mu^{\rm hlbl, NLO} = a_\mu^{\rm hlbl,LO}
\left ( 1 + \frac{\alpha_s}{\pi} C_F \frac{\Delta_1}{\Delta_0}
\right ), 
\label{eq4}
\ee
where $\Delta_0 = -19/16+3/2 \zeta(3)$ and 
\be
\begin{split}
& \Delta_1 = 
-\frac{473 \pi^2}{1080}\ln^22+\frac{52 \pi^2}{405}\ln^32
-\frac{42853 \pi^4}{259200}
\\
& +\frac{5771\pi^4}{32400}\ln2+\frac{473}{1080}\ln^42
 -\frac{52}{675}\ln^52   -\frac{8477}{2700}
\\
&  
+\frac{473}{45}a_4
+\frac{416}{45}a_5
+\frac{34727 \zeta_3}{2400}
 -\frac{23567 \zeta_5}{1440}. 
\label{eq8}
\end{split}
\ee
In Eq.~(\ref{eq8})
$a_{4,5}={\rm Li}_{4,5}(1/2)$ are the values of polylogarithmic functions and 
$\zeta_{3,5}$ are the values of Riemann zeta-function at the respective 
argument.  Numerically, $C_F \Delta_1/\Delta_0$ evaluates to 
$3.851$.

We take the ratio of Eqs.~(\ref{eq3},\ref{eq4}), equate 
it to the ratio of physical quantities and obtain 
\be
\label{eq5}
a_\mu^{\rm hlbl} = 
R^{\rm NLO}  \; a_\mu^{\rm hvp},
\ee
where $ R^{\rm NLO}= a_\mu^{\rm hlbl, NLO}/a_\mu^{\rm hvp, NLO}$.
The ratio of NLO contributions reads 
\be  
\begin{split}
& R^{\rm NLO} = 
\;f \left ( \frac{\alpha_s}{\pi} \right ) \;\frac{\alpha}{\pi} 
\left ( \frac{3}{2} \zeta_3 - \frac{19}{16} \right ) 
\frac{45 \langle Q_q^4 \rangle }{\langle Q_q^2 \rangle}.
\end{split}
\ee
The effects of  $\alpha_s$ corrections are absorbed into 
the function $f(\alpha_s)$, which  reflects 
the {\it relative} sensitivity of the hadronic vacuum polarization 
and hadronic light-by-light contributions to radiative corrections.
We find 
\be
f(x) = \frac{1+3.851\; x
}{
1+5.061 \; x
}.
\ee
This function depends only weakly on the strong coupling 
constant, changing  from $f(0) = 1$, 
to $f(1) = 0.8$, to $f(\infty) = 0.76$. Taking 
$a_\mu^{\rm hvp} = 6900 \times 10^{-11}$ and $\alpha_s$ in the 
interval $[0,..,\pi]$,  we find $a_\mu^{\rm hlbl}$ to be in the range 
$a_\mu = (118 - 148) \times 10^{-11}$, consistent with earlier 
model computations
\cite{Prades:2009tw}. The QCD effects tend to slightly 
decrease the hadronic light-by-light scattering contributions to the muon 
magnetic anomaly relative to hadronic light-by-light scattering, since 
$f(\alpha_s) < 1$.

In summary, we studied the prediction of the constituent quark model for the  
hadronic light-by-light scattering contribution to the muon magnetic anomaly,
including QCD radiative effects. While it is known that pions 
need to be explicitly introduced into the model to 
correctly describe the chiral limit 
of QCD,  we argued 
that, for the muon magnetic anomaly, this issue is numerically not 
important.  QCD radiative corrections to constituent quark loop 
diagrams provide a tool to diagnose the sensitivity of the vacuum 
polarization and the light-by-light scattering contributions 
to various aspects of QCD dynamics. 
We demonstrated that the 
ratio of hadronic vacuum polarization and hadronic light-by-light 
scattering contributions to the muon $g-2$ is remarkably stable against 
QCD radiative corrections.  Within our approach, we are not able 
to detect a large renormalization of the quark-photon 
vertex --  particular to hadronic light-by-light scattering contribution -- 
that is claimed to be responsible for the large enhancement of 
$a_\mu^{\rm hlbl}$ observed in Ref.~\cite{Goecke:2010if}.  We believe that 
our result indicates that the constituent quark 
model computation of the hadronic light-by-light contribution to  the 
muon magnetic anomaly is robust. 

{\bf Acknowledgments} 
We are grateful to A.~Czarnecki, F.~Petriello and A.~Vainshtein for  useful 
conversations. 
R.~B. thanks LBNL Theory Group for hospitality while 
part of this work was being performed.
This research is supported by the US DOE under contract
DE-AC02-06CH11357 and  by the NSF under grant PHY-0855365.

\end{document}